\documentstyle[aps,prl,multicol,epsf]{revtex}
\begin{document}
\draft
\title{Effect of Electron Correlation on the Bragg Reflection}
\author{Yasutami Takada and Manabu Kido}
\address{Institute for Solid State Physics, University of Tokyo, 
7-22-1 Roppongi, Minato-ku, Tokyo 106-8666, Japan}
\date{\today}
\maketitle
\begin{abstract}
We study the effect of correlation on the Bragg reflection in the 3D 
electron gas, the 1D Luttinger liquid, and the 1D Hubbard model in 
an alternating periodic potential at half-filling. In the last 
system, we suggest a Luttinger-liquid-type quasi-metallic state in 
the crossover region from the band insulator to the Mott insulator. 
We explain the appearance of this state in terms of the 
incompatibility of the Bragg reflection with the concept of 
Luttinger liquids. 
\end{abstract}
\begin{multicols}{2}
\narrowtext
One of the fundamental issues in solid state physics is to elucidate 
to what extent the concept of ``electronic band structure'' is 
relevant in a strongly-correlated system. 
In the free-electron gas, the band gap is formed by the quantum 
interference between an incident electron plane wave, 
$e^{i{\bf k \cdot r}}$, with a reflected one, 
$e^{i({\bf k+K}){\bf \cdot r}}$, where ${\bf K}$ is a 
reciprocal-lattice vector related to periodicity of $V$ 
an electron-lattice potential. 
In this sense, the quantum coherence leading to the Bragg reflection 
is a key to the band structure and thus we consider it quite 
important to study the correlation effect on the Bragg reflection. 

The study is conceptually simple in the Fermi liquid as represented 
by the three-dimensional electron gas (3DEG) at metallic densities. 
The point is that we should grasp the band-gap formation in terms of 
the Bragg reflection of a quasi-particle rather than a free 
electron, because quasi-particles or wave packets composed of a 
complicated combination of plane waves due to $U$ the 
electron-electron interaction manifest themselves 
in low-energy physics. 
In the one-dimensional Luttinger liquid (1DLL), however, 
the situation is not so simple \cite{CSCC}; 
we should ask even the very existence of the Bragg reflection 
and this constitutes one of the aims of this paper. 

Basically there exist two complementary approaches to the 
many-electron system in a crystal described by the Hamiltonian $H$ 
composed of $T$ the kinetic energy, $V$, and $U$. 
One is ``the band approach'' or $(T\!+\!V)+\!U$ in which the 
problem is reduced to the self-consistent determination of an 
effective one-body potential $\widetilde{V}$ by combining $V$ with 
the effect of $U$ in the Hartree-Fock-like mean-field approximation. 
If desired, the correlation effect due to $U$ (which is missed in 
$\widetilde{V}$) can be included by perturbation in $U$ with respect 
to the unpertubed part $T\!+\!\widetilde{V}$. 
Another is ``the correlated-electron approach'' or $(T\!+\!U)+\!V$ 
in which we consider a correlated-electron state defined by 
$T\!+\!U$ first and then include $V$ perturbatively. 
Note that these two approaches do not always provide the same 
conclusion, as the discussion on the Mott transition \cite{IFT} 
indicates. 
We can even imagine situations in which the competition 
between $V$ and $U$ brings about a state which neither 
approach describes well. 

In this paper, an example of such intriguing situations is shown 
on the basis of our finding that the Bragg reflection is usually 
incompatible with the concept of Luttinger liquids. 
More specifically we shall treat the 1D Hubbard model with an 
alternating periodic potential at half-filling, a system attracting 
much attention in relation to the neutral-ionic transition \cite{ABPA}, 
the ferroelectric perovskites \cite{EIT,RS}, and the crossover 
from the band insulator (BI) at $V \!\gg\! U$ to the Mott 
insulator (MI) at $V \!\ll\! U$ \cite{SGN,GST,TS}. 
We have made a study in the density-matrix renormalization group 
(DMRG) \cite{DMRG} to obtain the charge and spin gaps, $\Delta_c$ 
and $\Delta_s$, as well as the electron localization length $\lambda$ 
with the system size up to 400 sites (which is larger than any previous 
calculations by an order of magnitude). 
By comparing these exact results with approximate ones given in both 
the band and correlated-electron approaches and also by referring to 
the very recent result of Fabrizio {\it et al.} \cite{FGN}, we suggest 
the appearance of a Luttinger-liquid-type quasi-metallic state at 
the BI-MI crossover. 
Here by ``quasi-metallic'' we mean that it is not the same as 
``metallic'' because of nonzero $\Delta_c$, but the state is 
distinct from either BI or MI by such features as very small 
$\Delta_c$ and long $\lambda$. 

For better illustration of the Bragg reflection of a quasi-particle, 
let us start with 3DEG in a weak periodic potential $V$ for which 
$H$ is given in second quantization with the plane-wave basis as 
\cite{units} 
\begin{eqnarray}
\label{H-EG}
H\!&=& \! T\!+\!V\!+\!U\!= \!
\sum_{{\bf k}\sigma} \varepsilon_{{\bf k}} 
c_{{\bf k}\sigma}^{+}c_{{\bf k}\sigma}\!+\!
\sum_{{\bf K} \neq {\bf 0}}\!\sum_{{\bf k}\sigma}
V({\bf K})c_{{\bf k}\sigma}^{+}c_{{\bf k+K}\sigma}
\nonumber \\
&&+ {1 \over 2}\sum_{{\bf q} \neq {\bf 0}}\!
\sum_{{\bf k}\sigma}\!\sum_{{\bf k'}\sigma'}U({\bf q})
c_{{\bf k+q}\sigma}^{+}c_{{\bf k'-q}\sigma'}^{+}
c_{{\bf k'}\sigma'}c_{{\bf k}\sigma},
\end{eqnarray}
where $c_{{\bf k}\sigma}$ annihilates an electron specified by 
momentum ${\bf k}$ and spin $\sigma$, 
$\varepsilon_{{\bf k}}\!=\!{\bf k}^2/2m\!-\!\mu$ with $m$ 
the mass of a free electron and $\mu$ the chemical potential, 
$V({\bf K})$ the local electron-ion pseudopotential, 
and $U({\bf q})\!=\!4\pi e^2/{\bf q}^2$. 

Single-electron properties can be analyzed by the study of the 
thermal Green's function $G_{{\bf k}, {\bf k+K}}(i\omega _n)$ 
with $\omega _n$ a fermion Matsubara frequency, defined 
conventionally as
\begin{eqnarray}
\label{GF-EG}
G_{{\bf k}, {\bf k+K}}(i\omega _n)
\! \equiv \! - \! \int_0^{1/T} \! d\tau 
\langle T_{\tau}c_{{\bf k} \sigma}(\tau)c_{{\bf k+K} \sigma}^{+}
\rangle e^{i\omega_n \tau}. 
\end{eqnarray}
In the correlated-electron approach, we consider 3DEG without $V$ 
in the first step. 
Here $G_{{\bf k}, {\bf k+K}}(i\omega _n)$ is not zero only for 
${\bf K} \!=\! {\bf 0}$. 
Thus we simply write $G_{\bf k}^{{\rm EG}}(i\omega _n)$ and 
this function can be determined by the formally exact Dyson 
equation as shown diagrammatically in Fig.~1(a) with use of 
the vertex function 
$\Lambda ({\bf k'}i\omega_{n'},{\bf k}i\omega_{n})$ 
defined in Fig.~1(c). 
In the second step, we include $V$ in its lowest order 
by solving the equation in Fig.~1(b) to obtain 
$G_{{\bf k}, {\bf k+K}}(i\omega _n)$ in terms of 
$G_{\bf k}^{{\rm EG}}(i\omega _n)$ and 
$\Lambda ({\bf k'}i\omega_{n'},{\bf k}i\omega_{n})$. 

In order to make the physics as clear as possible, we shall be 
concerned only with the most interesting case in which $|{\bf K}|$ 
is equal to $2k_{\rm F}$ with $k_{\rm F}$ the Fermi wave number 
of 3DEG. 
Then, a band gap $\Delta$ opens at the Fermi level with 
the value of $2V({\bf K})$ in the noninteracting electron gas, 
while in the interacting system, by expanding the self-energy 
$\Sigma_{\bf k}^{{\rm EG}}(i\omega _n)$ $[=\! i\omega _n\!-\!
\varepsilon_{{\bf k}}\!-\!G_{\bf k}^{{\rm EG}}(i\omega _n)^{-1}]$ 
with respect to $\omega_n$ up to first order in line with weak $V$, 
we find easily that $\Delta$ is given exactly as 
$2z_{k_{\rm F}}\Lambda (k_{\rm F}0,-k_{\rm F}0)V({\bf K})$ 
\cite{HNW} with $G_{-k_{\rm F},k_{\rm F}}(i\omega _n)
\!=\!z_{k_{\rm F}}\Delta/2[(i\omega _n)^2\!-\!(\Delta/2)^2]$ where 
$z_{k_{\rm F}}$ is the quasi-particle renormalization factor at the 
Fermi surface. 
Thus the ratio, $z_{k_{\rm F}}\Lambda (k_{\rm F}0,-k_{\rm F}0)$, 
quantifies the effect of correlation on the Bragg reflection. 

We estimate this ratio quantitatively by employing the local-field 
correction $G_+(q)$ to represent the effect of the vertex function. 
Using the values of $G_+(q)$ as supplied by quantum Monte Carlo 
\cite{MCS}, we can calculate the ratio, together with its 
components, $z_{k_{\rm F}}$ and $\Lambda (k_{\rm F}0,-k_{\rm F}0)\!
=\!\{1\!+\!\alpha r_s [1\!-\!G_+(2k_{\rm F})]/2\pi\}^{-1}$, 
as a function of $r_s$ the interelectron distance in units of the 
Bohr radius with $\alpha\!=\!(4/9\pi)^{1/3}\!\approx \!0.521$. 
The results are shown in Fig.~1(d) in which we see that 
$\Lambda (k_{\rm F}0,-k_{\rm F}0)\! \approx \!1$. 
\begin{figure}[h]
\centerline{\epsfxsize=2.6truein \epsfbox{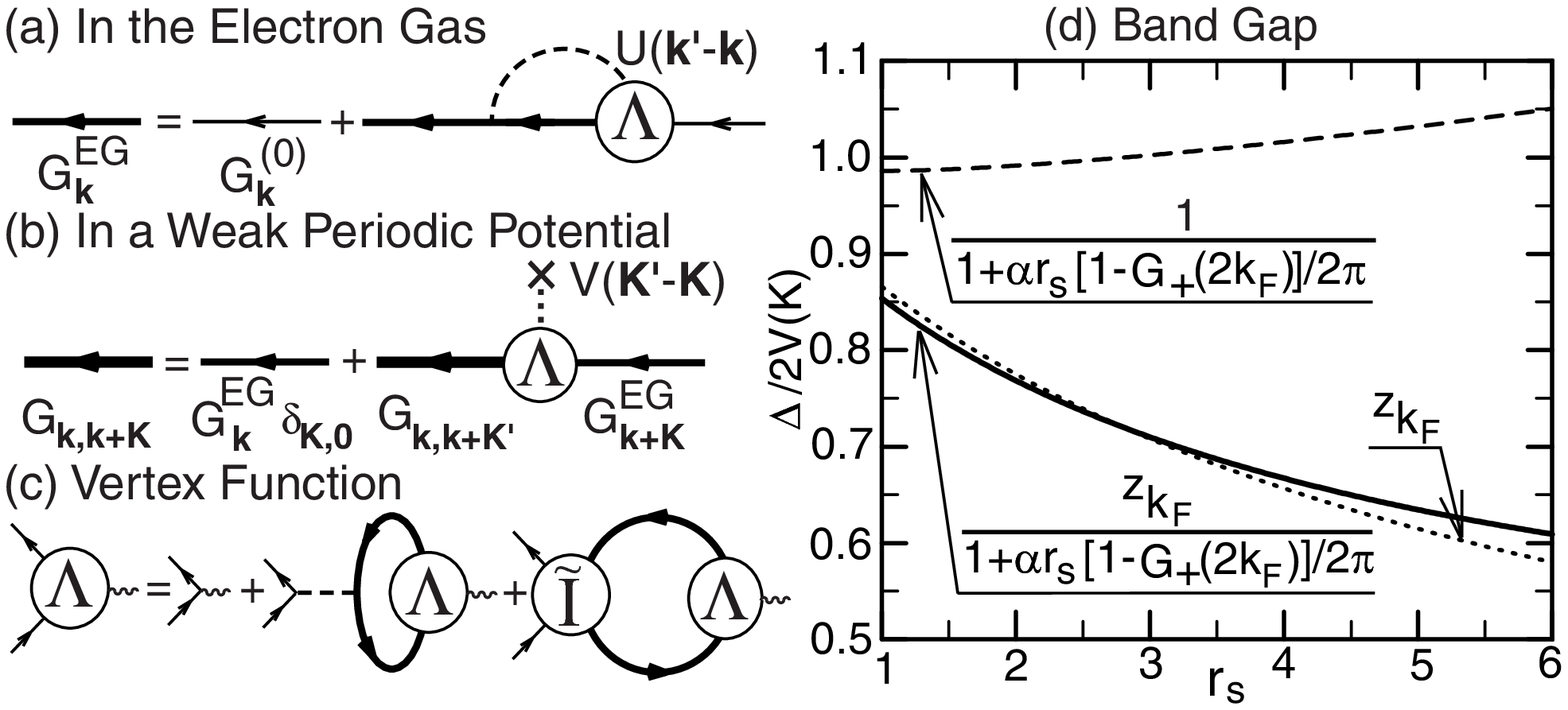} }
  \caption{Diagrammatic representation of 
    $G_{{\bf k}, {\bf k+K}}(i\omega _n)$ (a) without $V$ 
    and (b) with it in its lowest order. 
    The vertex function 
    $\Lambda ({\bf k'}i\omega_{n'},{\bf k}i\omega_{n})$ is defined 
    in (c) with use of the irreducible electron-hole effective 
    interaction $\widetilde{I}$. 
    In (d), effect of correlation on the Bragg reflection as 
    quantified by $\Delta/2V({\bf K}) = z_{k_{\rm F}}/\{1\!+\! 
    \alpha r_s [1\!-\!G_+(2k_{\rm F})]/2\pi\}$ is shown.} 
  \label{fig1}
\end{figure}

Deviation of the vertex function from unity accounts for the 
many-body effects on the electron-ion interaction $V_{\rm el-ion}$ 
and physically this should be small at metallic densities 
($1\!<\!r_s\!<\!6$) for $q\!=\!2k_{\rm F}$ corresponding 
to the interelectron distance in real space; 
$V_{\rm el-ion}$ is not modified much from the bare one in the 
neighborhood of an electron at this distance due to the absence of 
other electrons by correlation. 
Thus the ratio is essentially determined only by $z_{k_{\rm F}}$ 
or the weight of the coherent part, confirming a naively 
anticipated result that the Bragg reflection occurs only 
in the coherent part of a quasi-particle. 

The importance of $z_{k_{\rm F}}$ in the Bragg reflection prompts us 
to investigate 1DLL in which $z_{k_{\rm F}}$ vanishes. 
The Hamiltonian is basically the 1D version of Eq.~(\ref{H-EG}); 
$\varepsilon_{k}$ is linearized as $\varepsilon_{k}\!=
\!v_{\rm F}(k\!-\!k_{\rm F})\ [v_{\rm F}(-k\!-\!k_{\rm F})]$ 
for the right-\ [left-]moving branch in $T$ and coupling constants, 
$g_1$, $g_2$, and $g_4$, corresponding to the backward scattering, 
the forward scattering between the opposite branches, and the 
forward one within the same branch, respectively, are introduced 
in $U$ \cite{Solyom}. 
Using $a_{k\sigma}\ [b_{k\sigma}]$ the annihilation operator 
for an electron in the right- [left-]moving branch, 
we can write $V$ as
\begin{eqnarray}
\label{H-L}
V =  v \sum_{k\sigma}
(a_{k\sigma}^{+}b_{k-2k_{\rm F}\sigma} + 
b_{k-2k_{\rm F}\sigma}^{+}a_{k\sigma}), 
\end{eqnarray}
in which only the $K\!=\!2k_{\rm F}$ part for $V(K)$ is retained. 

The assertion we shall make is that $\Delta_c$ due to $V$ 
corresponding to $\Delta$ in the Fermi liquid vanishes in 1DLL. 
Our strategy to prove it is to find a criterion as to 
when $V$ turns out to be an irrelevant perturbation. 
Since the key quantity is $\langle a_{k_{\rm F}\sigma}^{+}
b_{-k_{\rm F}\sigma} \rangle$ or 
$T\sum_{\omega_n}G_{-k_{\rm F},k_{\rm F}}(i\omega_n)$, 
we evaluate the expectation value by treating $V$ as a linear 
perturbation to the system decribed by the Hamiltonian $T\!+\!U$. 
Then, the Kubo's formula provides us 
\begin{eqnarray}
\label{Kubo}
\langle a_{k_{\rm F}\sigma}^{+}b_{-k_{\rm F}\sigma} \rangle
= {v \over 2} \lim_{\omega \to 0}N(2k_{\rm F},\omega),
\end{eqnarray} 
where $N(q,\omega)$ is the charge-density response function. 
Thus the problem is reduced to evaluating $N(2k_{\rm F},\omega)$ 
in the $\omega \! \to \!0$ limit for the system of 
$T\!+\!U$ \cite{Fermi}. 
The behavior of this function is known well \cite{Solyom} and 
the result is $N(2k_{\rm F},\omega\to 0) \propto 
\omega^{\gamma_{\rho}-1}$ with $\gamma_{\rho}$, given by 
\begin{eqnarray}
\label{rho}
\gamma_{\rho}\! = \! \sqrt{{1+(g_4+g_1-2g_2)/2\pi v_{\rm F} \over 
1+(g_4-g_1+2g_2)/2\pi v_{\rm F}}},
\end{eqnarray} 
for $g_1 \ge 0$. 
This leads us to conclude that $\langle a_{k_{\rm F}\sigma}^{+}
b_{-k_{\rm F}\sigma} \rangle$ vanishes for $g_1\! >\!2g_2$ 
even in the presence of $V$. 
More generally, the expectation value is zero in the phases 
characterized by $N(2k_{\rm F},\omega\to 0)\!=\!0$, indicating 
the irrelevance of $V$. 
This irrelevance implies that $\Delta_c$ due to $V$ vanishes, 
because the system is the same as that with $v=0$.

Physically low-lying excitations in 1DLL are rigorously represented 
by sound waves with wavelengths much longer than $1/2k_{\rm F}$. 
Thus the effect of $V$ manifests itself after the average of $V$ 
over a distance longer than its periodicity, which is null and 
leads to the complete absence of the Bragg reflection. 
This statement ceases to be valid if $V$ is large enough 
to destroy the Luttinger-liquid state itself. 
In this sense, the concept of the Bragg reflection is incompatible 
with that of Luttinger liquids. 

So far no lattice periodicity is considered in $T\!+\!U$ and thus 
the concept of electron filling is irrelevant. 
Now we include it by treating a 1D system at half-filling 
on the lattice prescribed by $T\!+\!U$ with $V$ possessing 
periodicity of two lattice units. 
In site representation, $H$ is given by
\begin{eqnarray}
\label{H-Hubbard}
H\!&=& \! T\!+\!V\!+\!U\!= \!
-t\sum_{j \sigma} (c_{j\sigma}^{+}c_{j+1\sigma}\!+\!
c_{j+1\sigma}^{+}c_{j\sigma})
\nonumber \\
&&+v\sum_{j \sigma} (-1)^j c_{j\sigma}^{+}c_{j\sigma}
+u\sum_{j} c_{j\uparrow}^{+}c_{j\uparrow}
c_{j\downarrow}^{+}c_{j\downarrow}.
\end{eqnarray} 
For the study of competition between $V$ and $U$ in the 
whole region from $v\! \gg \! u$ to $v\! \ll \! u$, 
we implement DMRG to calculate 
$E(N_{\uparrow},N_{\downarrow})$ the ground-state energy 
with $N_{\sigma}$ the fixed number of $\sigma$-spin electrons 
in the $L$-site system under the open-boundary condition. 
At size $L$, the charge and spin gaps, 
$\Delta_c(L)$ and $\Delta_s(L)$, are given as 
\begin{eqnarray}
\label{charge}
\Delta_c(L)\!&=&\!E({L\over 2}\!+\!1,{L\over 2})
\!+\!E({L\over 2}\!-\!1,{L\over 2})
\!-\!2E({L\over 2},{L\over 2}),
\\
\label{spin}
\Delta_s(L)\!&=&\!E({L\over 2}\!+\!1,{L\over 2}\!-\!1)\!
-\!E({L\over 2},{L\over 2}).
\end{eqnarray} 
By using a finite-size scaling as $\Delta_i(L)\!=\!(\Delta_i^2\!
+\!A_i/L^2\!+\!B_i/L^3\!+\cdots)^{1/2}$ for $i\!=\!c$ or $s$ 
\cite{GST}, we extrapolate the data at $L\!=\!50$, 100, 200, and 400 
to obtain the values at $L\!=\!\infty$, $\Delta_c$ and $\Delta_s$. 
In Fig.~2, the results thus obtained are plotted as a function 
of $u$ at $v=0.5t$. 
As $u$ increases, both $\Delta_c$ and $\Delta_s$ decrease 
from $2v$ (the band gap at $u\!=\!0$) and become very small 
at around $u\!=\!2.6t$. 
With the further increase of $u$, $\Delta_c$ increases very rapidly, 
while $\Delta_s$ remains to be zero. 
In the inset of Fig.~2, the gaps near $u\!=\!2.6t$ are shown 
in detail. 
The same overall behavior of the gaps is seen for other 
values of $v$ of the order of $t$. 

\begin{figure}[h]
\centerline{\epsfxsize=2.0truein \epsfbox{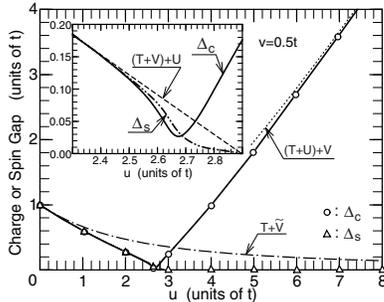} }
 \caption{Charge and spin gaps as a function of $u$ at $v=0.5t$. 
   Approximate values in the band and correlated-electron 
   approaches are shown by dashed and dotted curves, respectively, 
   while those in HF by the dotted-dashed curve. 
   The inset shows the results around $u\!=\!2.6t$ 
   in a magnified scale.} 
\end{figure}

Let us analyze these results by the comparison with those 
in both band and correlated-electron approaches. 
The former approach begins with the Hartree-Fock (HF) approximation 
which amounts to the one-body problem described by 
$T\!+\!\widetilde{V}$ where we define $\widetilde{V}$ in the same 
form of $V$ in Eq.~(\ref{H-Hubbard}) by replacing $v$ into 
$\tilde{v}$, determined through 
\begin{eqnarray}
\label{HF}
\tilde{v}\!=\! v-u\int_{-{\pi \over 2}}^{\pi \over 2}\! 
{dk \over 2\pi}
{\tilde{v} \over \sqrt{\tilde{v}^{2}\!+\!4t^{2}\cos^{2}k}}\ . 
\end{eqnarray}
In HF we obtain the band insulator (BI) brought about by the full 
Bragg reflection with $\Delta_{c} \!=\! \Delta_{s} \!=\! 2\tilde{v}$. 
Since $\tilde{v}$ is always positive, these gaps never vanish, 
which clearly contradicts the exact results for large $u$. 
Even for $u$ as small as $t$, they do not agree well with 
the exact ones, as seen by the dotted-dashed curve in Fig.~2. 

However, we achieve a surprisingly good improvement by including 
the correlation effect in second-order perturbation in $U$ 
with the unperturbed basis in $T\!+\!\widetilde{V}$, 
as shown by the dashed curve in the inset of Fig.~2. 
In fact, the exact gaps are reproduced quite accurately 
in this $(T\!+\!V)+\!U$ approach for $u$ from $0$ up to about 
$u_{c0}\! \equiv \! 2.45t$. 
Note that the correlation effect included in this way 
does not separate $\Delta_c$ from $\Delta_s$. 
Physical mechanism to reduce the gaps from those in HF is the same 
as explained in 3DEG, namely, the reduction of the the Bragg 
reflection on the conversion from a free electron to a 
quasi-particle. 
Thus we conclude that the state realized in the system 
for $u\!<\!u_{c0}$ is the correlated BI. 

For $u$ larger than about $u_{c2} \! \equiv \! 2.90t$, on the 
other hand, $\Delta_s$ vanishes, while $\Delta_c$ does not, 
indicating that the state is the MI. 
The effect of $v$ can be included in $\Delta_c$ 
in the $(T\!+\!U)+\!V$ approach; 
by examining the exact solution in $T\!+\!U$ \cite{LW}, 
we can deduce an expansion for $\Delta_{c}$ in $u^{-1}$ 
up to third order as 
\begin{eqnarray}
\label{Bethe}
\Delta_{c}  \approx  u \!-\! 2\sqrt{v^2\!+\!4t^2}
\!+\!8\ln 2 \ {t^2 \over u} {u^2\!+\!4v^2 \over u^2} 
\!-\! 6 \zeta (3) {t^4 \over u^3},
\end{eqnarray} 
with $\zeta (3)\!\approx \!1.202$. 
The above result is plotted by the dotted curve in Fig.~2 in which 
we see that the exact result is reproduced quite well for $u$ 
larger than $5t$. 

Now we need to clarify the nature of the state for $u$ 
at the BI-MI crossover, ranging from $u_{c0}$ to $u_{c2}$. 
For that purpose, we calculate $D_L$ a dimensionless localization 
parameter introduced by Resta and Sorella \cite{RS} as 
\begin{eqnarray}
\label{D}
D_L\!=\! -L \ln \Bigl |
\langle \Psi | \exp \Bigl (i{2\pi \over L} 
\sum_{j}x_j \Bigr ) |\Psi \rangle \Bigr |^2,
\end{eqnarray} 
with $\Psi$ the ground-state wavefunction at $N_{\uparrow}\!=\!
N_{\downarrow}\!=\!L/2$ under the open-boundary condition 
for the $L$-site system and $x_j$ the position operator at site $j$. 
Extrapolation of $D_L$ to the $L\!\to \!\infty$ limit gives 
the value $D$ which is related to $\lambda$ through 
$\lambda \!=\! \sqrt{D}/2\pi$ in units of the lattice spacing. 

The calculated results for both $D$ and $D_L$ at various $L$'s 
are shown as a function of $u$ in Fig.~3(a). 
For either $u\!<\!u_{c0}$ or $u\!>\!u_{c2}$, we see that $D_L$ 
converges to $D$ at $L$ as small as 100, while for $u$ in-between, 
even $L\!=\!400$ is not large enough for the convergence. 
In particular, $D$ seems to diverge for $u$ around $u_{c1} \! 
\equiv \! 2.65t$. 
Divergence in $D$, or equivalently that in $\lambda$, implies 
the appearance of a metallic state, indicating the vanishment of 
$\Delta_c$. 

As for $\Delta_c\!=\!0$ at $u\!=\!u_{c1}$ and the state for 
$u_{c1}\!<u\!<\!u_{c2}$, Fabrizio {\it et al.} \cite{FGN} have 
suggested a spontaneously dimerized insulating phase (SDI) 
by analytical arguments. 
In fact we find that $\lambda$ decreases quite rapidly to be less 
than twice the lattice spacing as $u$ increases from $u_{c1}$, 
implying an insulating behavior. 
Thus we conclude that an insulating phase, most likely SDI, is 
realized for $u_{c1}\!<\!u\!<\!u_{c2}$. 

\begin{figure}[t]
\centerline{\epsfxsize=3.5truein \epsfbox{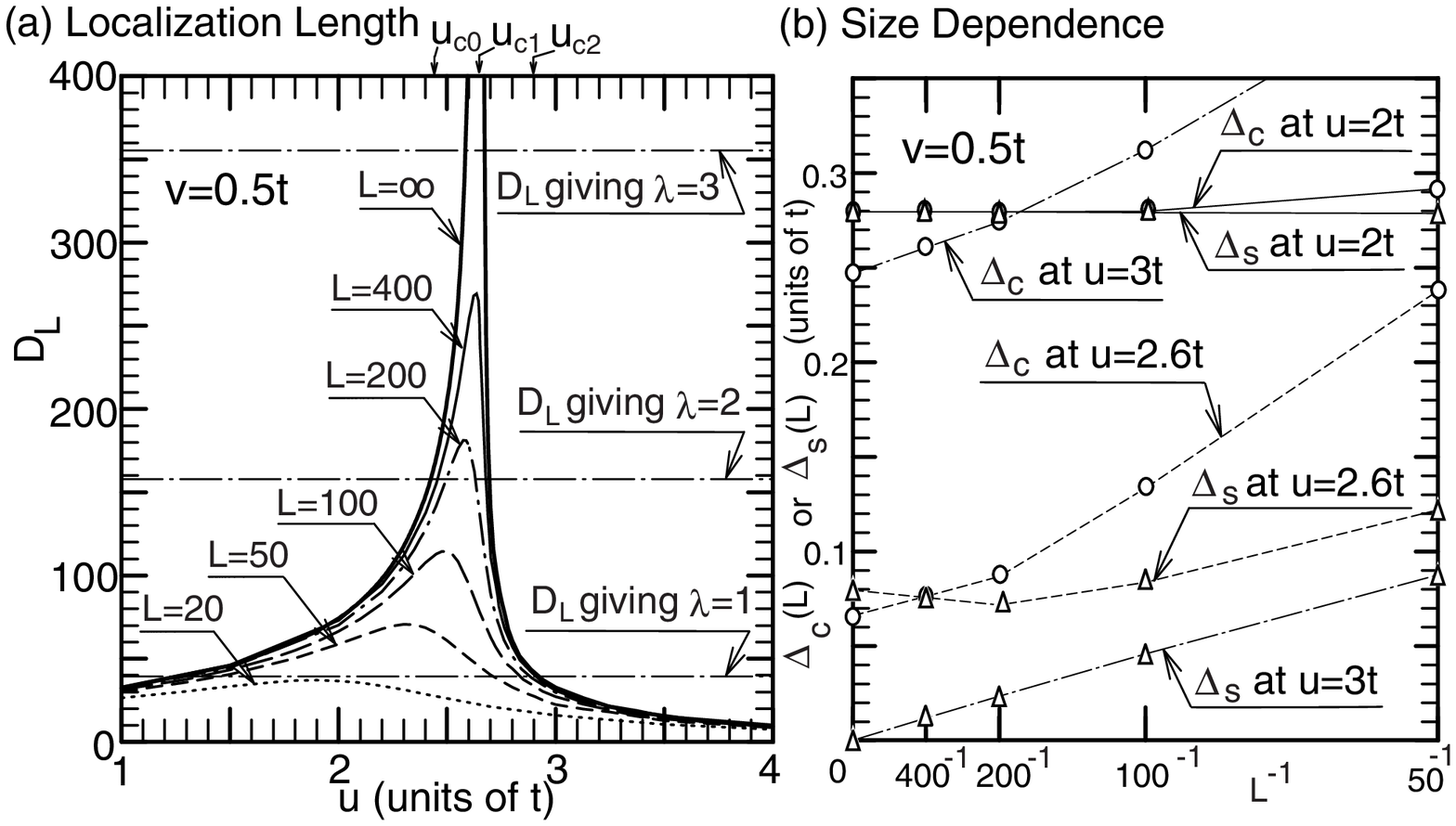} }
  \caption{(a) Dimensionless localization parameter $D_L$ as a 
     function of $u$ at $v\!=\!0.5t$ for various system sizes, 
     together with the value extrapolated to $L\!=\!\infty$. 
     The values of $D_L$ corresponding to $\lambda\!=\!1$, $2$, 
     and $3$ are also indicated.
    (b) Size dependnece of both $\Delta_c(L)$ and 
    $\Delta_s(L)$ at $u\!=\!2t$, $2.6t$, and $3t$. } 
\end{figure}

A remaining problem is that $\Delta_c$ never becomes zero 
in Fig.~2, although we now know that it should be zero at least 
at $u\!=\!u_{c1}$. 
Thus we reexamine the size dependence of both $\Delta_c(L)$ 
and $\Delta_s(L)$ carefully and the typical results are plotted 
in Fig.~3(b). 
Let us first analyze them in terms of $\Delta_c\!-\!\Delta_s$ or 
the spin-charge separation. 
Because of one dimensionality, one might assume that 
$\Delta_c\! \neq \! \Delta_s$ should be the case as long as 
$u\! \neq \! 0$, but this is not true; 
at $u\!=\!2t$ in the BI region, we find that both $\Delta_c$ and 
$\Delta_s$ coincide up to at least five digits (which exceeds 
numerical accuracy) at $L\!=\!200$ or larger. 
On the other hand, we know that $\Delta_c\!-\!\Delta_s
\!=\!-\Delta_s\!\neq \!0$ at $u\!=\!u_{c1}$. 
Therefore there definitely exists a value of $u$ at which 
$\Delta_c\!-\!\Delta_s$ begins to deviate from zero. 
We identify $u_{c0}$ as such a value and thus only for $u$ 
larger than $u_{c0}$ the 1DLL-like spin-charge separation occurs. 
In this sense, we consider that $u_{c0}$ gives a sharp phase 
boundary. 

In order to find more detailed features of the state at $u$ 
from $u_{c0}$ to $u_{c1}$, let us look at Fig.~3 again. 
As represented at $u\!=\!3t$ in Fig.~3(b), $\Delta_s(L)$ converges 
to zero very nicely with the increase of $L$ in MI. 
Similar convergence of both charge and spin gaps is obtained 
in SDI as well. 
However, a distinct behavior is seen in the gaps 
for $u_{c0}\!<\!u\!<\!u_{c1}$ as illustrated at $u\!=\!2.6t$; 
both $\Delta_c(L)$ and $\Delta_s(L)$ at $L\!=\!400$ seem to be 
larger than those extrapolated from the data at smaller $L$, 
implying that our data for the gaps are not accurate enough 
for these $u$'s. 
This inaccuracy should be due to the large $\lambda$ which is always 
longer than the two lattice units for these $u$'s as indicated 
in Fig.~3(a). 
This difficulty in obtaining exact values for the gaps in this region 
can be overcome only by a more accurate calculation of energies 
at much larger $L$. 
Such a calculation is not feasible at present, but we can safely 
conclude even at the present time that $\Delta_c$ is very small, 
i.e., much smaller than $0.1t$ at most of the values of $u$ 
in this phase. 
The nature of small $\Delta_c$ and long $\lambda$ suggests 
us that a Luttinger-liquid-type quasi-metallic phase appears for 
$u_{c0}\!<u\!<\!u_{c1}$. 
Incidentally, each electron in this phase feels $V$ with the spatial 
average over $\lambda$ which is longer than the periodicity of $V$. 
This implies that the effect of $V$ on the electrons is very small 
and thus this should be the reason why the feature of BI is lost 
in the state for $u_{c0}\!<u\!<\!u_{c1}$. 
Here again we find that the Bragg reflection, a crucial concept 
to define BI, is incompatible with the Luttinger-liquid feature. 

Finally we note that both this phase and SDI deserve special 
attention, because they are not anticipated in both band and 
correlated-electron approaches; their existence is entirely 
due to the competition of $V$ and $U$.

In conclusion, we have investigated the effect of electron 
correlation on the Bragg reflection in a variety of situations 
in various approaches. 
We have found the incompatibility of the Bragg reflection with 
the Luttinger liquid, based on which we have suggested a 
Luttinger-liquid-type quasi-metallic state at the crossover from BI 
to MI via SDI. 

Y.T. is supported by the Grant-in-Aid for Scientific Research (C) 
from the Ministry of Education, Science, Sports, and Culture of Japan. 


\end{multicols}

\begin{references}
\bibitem{Fermi} In the Fermi liquid, $N(2k_{\rm F},\omega) 
\propto z_{k_{\rm F}}^2/\omega$ and it diverges. 
By replacing $\omega$ by $\Delta$ in $N(2k_{\rm F},\omega)$ 
in the limit of $\omega \to 0$, we obtain 
$\langle c_{k_{\rm F}\sigma}^{+}c_{-k_{\rm F}\sigma} 
\rangle \!=\!-z_{k_{\rm F}}/2$.
\end{references}
\end{document}